\documentclass[preprint,nofootinbib,amsmath,amssymb,amsfonts,aps,prd,groupedaddress]{revtex4-1}

\usepackage{amssymb,amsfonts}
\usepackage{setspace} \usepackage{graphicx} \usepackage{subfigure}
\usepackage[bookmarks=true, pdfstartview={FitH}, bookmarksopen=true,
bookmarksopenlevel=1, linktoc=page]{hyperref} \usepackage{amsthm}
\usepackage{mathrsfs} 
\usepackage{tikz}
\usetikzlibrary{patterns}

\newcommand{\be}{\begin{equation}} \newcommand{\ee}{\end{equation}}
\newcommand{\ba}{\begin{equation}\begin{aligned}}
\newcommand{\ea}{\end{aligned}\end{equation}}

\DeclareMathAlphabet{\mathpzc}{OT1}{pzc}{m}{it}

\usepackage{amsthm} \let\oldproofname=\proofname
\renewcommand{\proofname}{\rm\bf{\oldproofname}:}

\usepackage{mathrsfs}

\begin{document}

\title{Information Loss}

\author{William G. Unruh}
\email{unruh@physics.ubc.ca}
\affiliation{Department of Physics and Astronomy, University of British
Columbia, Vancouver, BC, Canada V6T 1Z1}

\author{Robert M. Wald}
\email{rmwa@uchicago.edu}
\affiliation{Enrico Fermi Institute and Department of Physics, The University
of Chicago, 5640 South Ellis Avenue, Chicago, Illinois 60637, USA}

\begin{abstract}

The complete gravitational collapse of a body in general relativity will result
in the formation of a black hole. Although the black hole is classically stable,
quantum particle creation processes will result in the emission of Hawking
radiation to infinity and corresponding mass loss of the black hole, eventually
resulting in the complete evaporation of the black hole. Semiclassical arguments
strongly suggest that, in the process of black hole formation and evaporation, a
pure quantum state will evolve to a mixed state, i.e., there will be
``information loss.'' There has been considerable controversy over this issue
for more than 40 years. In this review, we present the arguments in favor of
information loss, and analyze some of the counter-arguments and alternative
possibilities.

\end{abstract}

\maketitle

\section{Introduction}

Ever since Hawking's discovery \cite{hawk} that black holes radiate thermally and
should therefore ``evaporate'' completely, battles have raged about the final
quantum state resulting from this process.  Semiclassical arguments clearly
indicate that the final state should be mixed because of the entanglement of the
Hawking radiation with degrees of freedom inside the black hole and the ultimate
disappearance of these degrees of freedom (with respect to our universe, at
least) when the black hole evaporates. It is very difficult to find a flaw in
these semiclassical arguments that would invalidate this conclusion. On the
other hand, it has been argued on a variety of grounds that such pure state to
mixed state evolution would violate fundamental principles of physics and/or
other cherished beliefs. This conflict has thus given rise to the so-called {\it
black hole information loss paradox}. 

Our belief is that information is lost into black holes, i.e., in the process of
black hole formation and evaporation, a pure state will evolve to a mixed
state\footnote{However, as discussed at the end of section \ref{alt} below, we
hold open the possibiity of a pure final state as a result of entanglement of
the Hawking radiation with the vacuum left behind after black hole
evaporation.}. In classical general relativity, the singularity inside the black hole acts as
a sink for any ingoing radiation, which removes it from the universe. In quantum gravity,
the classical singularity will undoubtedly be replaced by
some other, weirder structure, but it should still act as a sink. 

The aim of this note is to review the arguments in favor of
information loss and to argue that such pure to mixed evolution does not violate
any fundamental principles of physics. We begin in section \ref{entan} with a
review of entanglement in quantum mechanics and quantum field theory. As
explained in section \ref{infoloss}, this gives rise to the semiclassical
arguments in favor of information loss. Possible alternatives to this
semiclassical picture are discussed in section \ref{alt}. Arguments against
information loss are analyzed in section \ref{noinfo}. Our conclusions are given
in section \ref{con}.

\section{Entanglement} \label{entan}

Entanglement is a ubiquitous feature of quantum mechanics, a feature which
Schrodinger \cite{schr} referred to as ``{\it the} characteristic trait of
quantum mechanics, the one that enforces its entire departure from classical
lines of thought.''  Suppose that we have two quantum mechanical systems, the
states of which are individually represented by Hilbert spaces ${\mathcal H}_1$
and ${\mathcal H}_2$, respectively. Then, by the rules of quantum mechanics, the
possible states of the joint system are represented by the tensor product,
${\mathcal H}_1 \otimes {\mathcal H}_2$, of these Hilbert spaces. This tensor
product Hilbert space is the (closure of the) span of product states of the form
\be | \Psi_1 \rangle \otimes  | \Psi_2 \rangle \, , \label{prod} \ee where $|
\Psi_1 \rangle \in {\mathcal H}_1$ and $| \Psi_2 \rangle \in {\mathcal H}_2$.
Thus, a general state in the tensor product Hilbert space is of the form \be |
\Psi \rangle = \sum_i c_i | \Psi_{1i} \rangle \otimes  | \Psi_{2i} \rangle \, .
\label{ent} \ee We can encode the information about this state relevant to
observations concerning only the first system via a density matrix $\rho_1 :
{\mathcal H}_1 \to {\mathcal H}_1$ defined by \be \rho_1 = \sum_{ij} c_i c_j^*
\langle \Psi_{2j}| \Psi_{2i} \rangle |\Psi_{1i} \rangle \langle \Psi_{1j}| \, .\ee A
normalized state $| \Psi \rangle$ can be written in the form of a product state,
eq.~\eqref{prod}, if and only if $\rho_1^2 = \rho_1$, in which case system 1 (and
system 2) is said to be in a {\it pure state}. For such a product state, system
1 behaves just as if it is the state $| \Psi_1 \rangle$ and the presence of
system 2 is irrelevant to measurements of system 1; similarly, system 2 behaves
as if it is in state $| \Psi_2 \rangle$ and the presence of system 1 is
irrelevant. However, generically a state $| \Psi \rangle$ of the form
\eqref{ent} cannot be written as a product state \eqref{prod}, in which case
systems 1 and 2 are said to be {\it entangled}, and the state of system 1 (as
well as system 2) on its own is said to be {\it mixed}. For entangled systems, there are
nontrivial correlations between the outcomes of measurements made on the two
systems.  In particular, if and only if the state is entangled, one can find an
observable ${\mathcal O}_1$ for system 1 and an observable ${\mathcal O}_2$ for
system 2 such that
 \be \langle \Psi | {\mathcal O}_1 \otimes {\mathcal O}_2 |
\Psi \rangle \neq \langle \Psi | {\mathcal O}_1 | \Psi \rangle \langle \Psi |
{\mathcal O}_2 | \Psi \rangle \, .  \label{entobs} \ee
Entanglement is
ubiquitous because even if systems 1 and 2 are initially in a product state
\eqref{prod}, they will generically evolve to an entangled state in the presence
of any nontrivial interaction. They will then remain entangled even after they
cease interacting.

Although entanglement is a ubiquitous feature of quantum mechanics in general, it is an
essential feature of quantum field theory. In the case of quantum field theory,
the full system consists of the quantum field observables over all of spacetime,
or, equivalently---assuming deterministic evolution---the quantum field
observables in a neighborhood of any Cauchy surface, $\Sigma$. We can divide the
quantum field system into two subsystems by dividing the Cauchy surface into
disjoint open regions $\Sigma_1$, $\Sigma_2$, with common boundary $S$, such
that $\Sigma_1 \cup \Sigma_2 \cup S =\Sigma$.  Let ${\mathcal U}_1$ and
${\mathcal U}_2$ be globally hyperbolic regions with Cauchy surfaces $\Sigma_1$
and $\Sigma_2$, respectively, as illustrated in Fig. 1. 
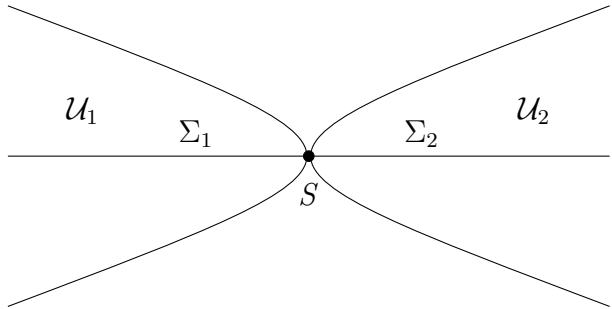
\begin{figure}[ht]
\centering
\begin{tikzpicture}
\draw (-4,0)  -- (4, 0);
\filldraw [black] (0, 0) circle (2pt);
\draw (4,-2) .. controls (-1.3,0) .. (4,2);
\draw (-4,-2) .. controls (1.3,0) .. (-4,2);
\node at (0,-.5) {\textit{S}}; 
\node at (1.5,.3) {\textit{$\Sigma_2$}}; 
\node at (-1.5,.3) {\textit{$\Sigma_1$}}; 
\node at (-3,.6) {\textit{${\mathcal U}_1$}}; 
\node at (3,.6) {\textit{${\mathcal U}_2$}};
\end{tikzpicture}
\caption{A spacetime diagram showing the adjoining regions ${\mathcal U}_1$ and ${\mathcal U}_2$ described in the text. In any physically reasonable state, the quantum field observables in ${\mathcal U}_1$ are highly entangled with the quantum field observables in ${\mathcal U}_2$.}
\end{figure}
We take system 1 to be
the field observables in ${\mathcal U}_1$ and system 2 to be the field
observables in ${\mathcal U}_2$.  Then, for any relativistic quantum field theory, in any physically acceptable state, these
two systems will be entangled.  To see this 
explicitly, let $x \in S$ and let both $x_1
\in \Sigma_1$ and $x_2 \in \Sigma_2$ approach $x$. For the particular case of a free
scalar field, $\phi$, the leading order
behavior for $x_1, x_2$ near $x$ for any physically acceptable (Hadamard) state,
$\Psi$, is given by
\be \langle \Psi | \phi(x_1) \phi(x_2) | \Psi \rangle \sim \frac{U(x_1,x_2)}{(2
\pi)^2 \sigma(x_1, x_2)} \, , \label{2pt}\ee 
where $U(x_1, x_2)$ is smooth and $\sigma(x_1, x_2)$ is the squared geodesic
distance between $x_1$ and $x_2$. Thus, the right side of \eqref{2pt} diverges as $x_1,
x_2 \to x$. On the other hand, for any physically acceptable state, as $x_1, x_2 \to x$, we have
\be \langle \Psi | \phi(x_1) | \Psi \rangle
\langle \Psi | \phi(x_2) | \Psi \rangle \to \left[\langle \Psi | \phi(x) | \Psi
\rangle\right]^2 \label{1pt} \ee which is not
divergent. 
Thus, the left sides of \eqref{2pt} and \eqref{1pt} cannot be equal, and there is entanglement.
Similar behavior hold for all fields (including
interacting fields) for all physically acceptable states. Thus, \eqref{entobs}
always holds, and there always is entanglement between systems 1 and 2 in any
physically acceptable state.

An important example of the above phenomenon is the case of Minkowski spacetime,
with ${\mathcal U}_1$ and ${\mathcal U}_2$ taken to be the two Rindler wedges
and $| \Psi \rangle$ taken to be the Minkowski vacuum state.  The above
discussion shows that there will be entanglement between the field observables
in the left and right Rindler wedges. This accounts for why the family of
observers in one of the Rindler wedges that follows orbits of Lorentz boosts
sees a mixed state when the entire quantum field system is in the
(pure) Minkowski vacuum. The Minkowski vacuum is entangled between the two wedges. That is is also
a thermal state is a consequence of the particular nature of that entanglement.

 Our discussion above makes it clear that the
entanglement between the field in two such causally complementary regions {\it
always} occurs in quantum field theory, no matter what the spacetime or the
(physically acceptable) state.

\section{Information Loss} \label{infoloss}

A spacetime of considerable interest is one in which a black hole forms by
gravitational collapse. Just as in the general case above, at any ``moment of
time,'' $\Sigma_0$ after the black hole has formed, in any regular state (such as the state 
obtained by evolution from any physically 
reasonable initial state before the black hole formed), there
will be entanglement between the state of quantum field observables inside and
outside of the horizon of the back hole. This entanglement is intimately related to the nature
of the Hawking radiation emitted to infinity, allowing it to be in mixed state, and thermal.\footnote{We use the word ``thermal'' here in the sense that if one places a Schwarzschild black hole in a thermal bath at the Hawking temperature, then detailed balance will hold and the system will be in (possibly unstable) equilibrium. If the black hole is radiating into empty spacetime, then since some of the Hawking radiation is reflected back into the black hole by the curvature and angular momentum barriers around 
the black hole, the population of modes as seen by a distant observer will be reduced by this reflection. Similar behavior will occur for any other ``black body of finite size.''} 

The mode inside the black hole that is
maximally correlated with a Hawking radiation mode outside is called the partner state (see \cite{wald1}, \cite{hsu}). If we trace the behavior of a Hawking mode and its partner mode backward in time in a free quantum field theory, we find that the entanglement of the Hawking radiation
with the state of the quantum field inside the black hole arises from entanglement across the event horizon at very (i.e., transplanckian) short distance
scales at early times. One would not expect the free quantum field theory description of this entanglement to be valid in this regime. However, the study of analog black holes shows that 
the creation of thermal radiation does not rely on such transplanckian behavior: In analog models where the dispersion relations are modified at very short wavelengths, there is no change to
the prediction of Hawking radiation---and
its entanglement across the horizon---at wavelengths long compared with this modification scale
\cite{unr95}, \cite{cj}.
Furthermore, the argument of Fredenhagen and Haag \cite{fh}, \cite{hw} shows that Hawking radiation for black holes
can be derived as a consequence of the Hadamard behavior of the state near the horizon at relatively late times, without the necessity to evolve backward all the way into a transplanckian regime.

It should be noted that there may also be considerable
entanglement between the quantum field observables inside and outside of the
black hole produced by much more mundane processes that do not involve short
distance phenomena. For example, one can consider two physically separated
matter systems that (due, e.g., to prior interactions) are highly entangled with
each other. One can drop one of these matter systems into the black hole (or
make it part of the matter system that collapses to form the black hole in the
first place) and keep the the second matter system at a safe distance outside of the black
hole. There will then be considerable entanglement between the observables
inside and outside the black hole above and beyond what is predicted to occur
via the Hawking effect. If one wishes to eliminate all entanglement between observables inside
and outside of the black holes, one would have to eliminate entanglement arising from this
means as well.

As indicated by our discussion above, the presence of entanglement between the
quantum field observables inside and outside of a black hole at some relatively early time,
$\Sigma_0$, after black hole formation (as
shown in Fig.~2 below) is completely in accord with normal behavior in quantum field theory.
Assuming that one started with a pure state before the black hole was formed,
the full state of the quantum field at time $\Sigma_0$ is still pure, even
though the state of the quantum field outside of the black hole is mixed.
However, when back reaction effects of the quantum field on the black hole are
taken into account, the entanglement between the quantum field observables
inside and outside of the black hole gives rise to a ``{\it loss of
information}'' as follows: The Hawking radiation carries a flux of positive
energy to infinity, so there must be a corresponding flux of negative energy
going into the black hole\footnote{Quantum fields can have locally negative
energy densities and fluxes even if the corresponding classical fields always
satisfy positive energy conditions.}. This flux of negative energy can be viewed as 
originating outside the horizon---very crudely
at a scale of about a Schwarzschild radius outside---and is thus not directly tied to the entanglement.
This negative energy flux will reduce the
mass and area of the black hole, in full accord with conservation of total
energy and the generalized second law of thermodynamics. However, the flux of
Hawking radiation to infinity---and the corresponding flux of negative energy
into the black hole---increases as the black hole gets smaller. As a result, one
predicts that the black hole should ``evaporate'' completely in a finite time,
resulting in a spacetime as depicted in Fig. 2. 
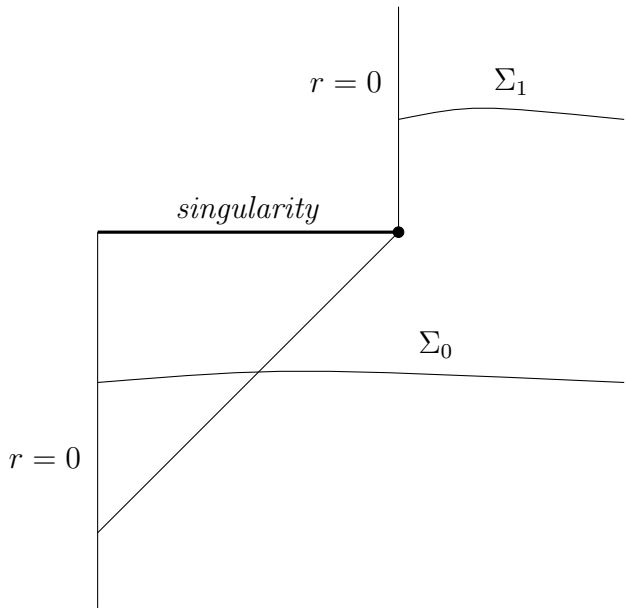
\begin{figure}[ht]
\centering
\begin{tikzpicture}
\draw (-2,-4)  -- (-2, 1);
\draw (-2,-3)  -- (2, 1);
\draw [very thick] (-2,1)  -- (2, 1);
\draw (2,1)  -- (2, 4);
\filldraw [black] (2, 1) circle (2pt);
\draw (-2,-1) .. controls (.5,-.8) .. (5,-1);
\draw (2,2.5) .. controls (3,2.7) .. (5,2.5);
\node at (-2.7,-2) {\textit{$r=0$}}; 
\node at (1.3,3) {\textit{$r=0$}};
\node at (0,1.3) {\textit{singularity}};
\node at (2.5,-.5) {\textit{$\Sigma_0$}};
\node at (3.5,3) {\textit{$\Sigma_1$}};
\end{tikzpicture}
\caption{A spacetime diagram of a black hole that evaporates. At time $\Sigma_0$ the observables 
inside and outside the black hole are entangled, just as in Fig.~1. The observables at time $\Sigma_1$ remain entangled with the observables inside the black hole---even though the black hole has evaporated.}
\end{figure}
The key point is that for a
quantum field propagating in the classical spacetime of Fig. 2, the state of the
quantum field on a late time surface $\Sigma_1$ after black hole evaporation
will remain entangled
with the quantum field observables inside the past black hole---even though the black
hole has evaporated completely and no longer exists!  Although the system was
initially in a pure state, the entire system at time $\Sigma_1$ is described by
a mixed state, so evolution from a pure state to a mixed state has occurred.
Furthermore, a complete knowledge of the state at time $\Sigma_1$ is
insufficient to determine the initial state. In this sense, information has been
lost to the black hole! We note that this argument holds whether the field inside has fallen off the edge of the
spacetime into a central singularity as occurs for a classical black hole, or, e.g., has gone into a baby universe created
by some quantum gravity effects near what would have been a singularity in the classical treatment. 

We emphasize that throughout the process of black hole formation and evaporation, the quantum field 
obeys deterministic, causal equations of motion. At no stage, except perhaps at the singularity, is there any breakdown 
of any known laws of physics, including local unitarity. The loss of information has arisen solely because the inside of the black hole
has acted as a sink as seen from outside for some of the degrees of freedom of the quantum field,
just as infinity is a sink for degrees of
freedom of massless fields escaping from any finite region of the spacetime. We will discuss this further in section V.A below.

\section{Alternatives to Information Loss} \label{alt}

The loss of information described in the previous paragraph is sufficiently
startling that it worthwhile to examine the assumptions on which it is based,
and to consider the ways in which these assumptions might be wrong. As described above, information
loss follows directly from applying the local evolutionary laws of quantum field
theory in a classical curved spacetime of the sort depicted in Fig.~2. However,
this analysis treats gravity classically, so this semiclassical analysis can at
best be an approximation to the true way that nature behaves. Nevertheless, in
order that the conclusion of information loss be invalidated, the semiclassical
approximation must fail in a significant way during some stage of the evolution.
It is useful to classify the possible ways in which it might fail into the following 4
categories\footnote{The recent work of Hawking, Perry, and Strominger \cite{hps} proposes
that black holes can have ``soft hair'' due to BMS-like charges. However, this work does not
propose any mechanism by which the presence of ``soft hair'' would enable the entanglement between quantum observables outside and inside the 
black hole can be avoided so that the final state resulting from black hole evaporation is pure. 
Since no concrete mechanism has been proposed, we cannot categorize this work in terms of our classification scheme below, but any such mechanism presumably would be subject to the criticisms of possibilities III and IV below.}:

\medskip

\noindent {\bf Possibility I: No black hole ever forms (fuzzballs):} Perhaps a
black hole never actually forms in the collapse, but rather---perhaps due to a
quantum tunneling phenomenon---some other structure without an event horizon,
such as a fuzzball \cite{math}, is formed. In that case, there would be no loss of
information to black holes simply because there are no black holes.

In our view, this is an extremely radical proposal. Classically, sufficiently
massive black holes form at arbitrarily low energy densities and curvatures, so
semiclassical general relativity and/or quantum field theory would have to
drastically fail in a regime where, {\it a priori}, one would expect these
descriptions to be extremely good. If one imagines a massive shell collapsing
radially inward at nearly the speed of light, we find it hard to imagine how it
could halt its inward momentum at just the right moment to form a fuzzball or
other structure without violating local conservation of momentum, causality, and
other basic principles of physics in a low curvature regime. The effective stresses
(as expressed by the Einstein curvatures of the effective spacetime metric) would have be absurdly
large---vastly exceeding the effective energy densities. Since horizons (true or apparent)
are global, not local, structures, the collapsing matter also would have to have non-local knowledge in order to start
behaving in this radical manner in order to avoid having a horizon form. 

Similar remarks apply to proposals (see, e.g., \cite{mh})
for huge local quantum 
backreaction effects 
to occur at the time when, classically, a black hole would form, thereby preventing the formation
of the black hole.

\medskip \noindent {\bf Possibility II: Major departures from semiclassical
theory occur during evaporation (firewalls):} Perhaps a black hole forms in the
expected manner but there are major departures from semiclassical theory during
the evaporation process, in such a way that there is greatly diminished
entanglement (or no entanglement at all) between the between the state of
quantum field observables inside and outside of the black hole. 

This is also a very radical proposal, since the destruction of entanglement
between the inside and outside of the black hole during evaporation requires a
breakdown of quantum field theory in an arbitrarily low curvature regime. In
particular, significantly diminished entanglement across the event horizon would
require the quantum field to be singular on the horizon---converting it to a
``firewall'' \cite{amps}. As in the case of fuzzball-type proposals,
the laws governing the creation of firewalls would have to
be drastically non-local/acausal in nature. In particular, it is difficult for us to see
how the firewall proposal could be made to work unless the firewall is located at exactly
the event horizon, which requires knowing the entire future history of the spacetime.
Even if the teleological requirement that the firewall be formed on the true event
horizon could be evaded, all other notions of ``horizon" (such as ``apparent horizon'') are
non-local in space, so acausal rules for the formation of firewalls would still be required.
Furthermore, if there is entanglement between matter that formed the black hole and matter that never fell in as mentioned in section III above, it would seem that one would need major violations of causal evolution unrelated to Hawking radiation to destroy this entanglement.

Instead of having drastic local departures from semiclassical theory at firewalls very near the 
horizon, one could hypothesize
departures that are less singular \cite{gid} but involve significant violations of causality at low energies. However, if this were to occur, 
it is difficult for us to see how the laws of nature would forbid
similarly large violations of causality when black holes are not present.

\medskip \noindent {\bf Possibility III: The black hole does not evaporate
completely (remnants):} Perhaps the evaporation process shuts off by the time
the black hole has evaporated down to the Planck scale---when quantum gravity
effects become dominant. The resulting ``remnant'' continues to contain all of
the ``information'' that went into the black hole (i.e., the remnant is highly
entangled with the exterior of the remnant), in such a way that the joint state
of the remnant and exterior is pure.

This is not a radical proposal, since it requires a breakdown of the
semi-classical picture only near the Planck scale, where it is expected to break
down. However, a crucial issue is whether such Planck scale remnants can interact with the
outside world. If they cannot, then 
it is not clear what ``good" the remnants do, since the
``information,'' although still present, is inaccessible, and, in practice, the final state 
will still be mixed. On the other hand, if the remnants can interact with the outside
world, then serious thermodynamic problems arise:
The initial black hole could have been
arbitrarily large, or have been fed by matter for an arbitrary long time,
so the remnants would need to have arbitrarily many states in
order to be entangled with all of the Hawking radiation emitted during the
evaporation process. Thus, if remnants could partake in the thermodynamics of the 
outside spacetime, they should be (infinitely) entropically favored over all
other types of matter, which suggests that they should be spontaneously produced at an 
arbitrarily high rate.

An additional problem with having remnants storing all of the ``information'' needed to
restore purity with Hawking radiation is that the Bekenstein-Hawking formula strongly
suggests that a Planck-sized remnant should have entropy $\sim 1$---i.e., it
should have $\sim 1$ possible state---rather than arbitrarily many.

\medskip \noindent {\bf Possibility IV: The information comes out in a final
burst:} Perhaps the evaporation process proceeds as in the semiclassical
analysis until the black hole reaches the Planck scale. However, perhaps all of
the information that had been stored within the black hole then emerges in a
final burst, so the final state is pure.

As in the case of remnants, this is not a radical proposal, since it requires a
breakdown of the semi-classical picture only near the Planck scale, where it is
expected to break down. However, at first glance, this proposal may seem absurd,
since it requires an arbitrarily large amount of ``information'' to be released
from an object of Planck mass and size. Clearly, this would not be possible if
the large amounts of ``information'' in the sense relevant here were required to
carry off correspondingly large amounts of energy, e.g., if it were necessary
for the black hole to emit some sort of burst of ``ordinary particles''
entangled with the Hawking radiation to carry this information. However, this is
not necessarily the case because ``vaccum fluctuations'' can contain arbitrarily
large amounts of ``information.'' Indeed, recently, Hotta, SchŸutzhold, and
Unruh \cite{hsu} have considered the model of a mirror in $(1+1)$-dimensions that
accelerates in such a manner as to emit Hawking-like radiation and then becomes
inertial. This is clearly a ``unitary" process, but during the accelerating phase the 
quantum field emission from the mirror is thermal in analogy with the emission from a black hole,
i.e.,  the analog of Hawking radiation is thermal, but the full state of the quantum
field must be pure at all times. The purity of the full state can be understood
as a consequence of entanglement of the Hawking radiation with ``partner
particles'' that are present outside of the mirror and eventually ``bounce off''
the mirror after it becomes inertial. However, these ``partner particles'' are
locally indistinguishable from vacuum fluctuations! The state of the quantum
field as seen by an observer at late retarded times (after the mirror has become
inertial) would be indistinguishable from the ordinary vacuum state. But the
vacuum fluctuations seen by such an observer would be correlated with the
Hawking radiation in such a way as to produce a pure state. 

It would seem more difficult to have analogous behavior in the black hole case,
where the highly non-classical ``burst'' region should be of Planck scale in
time and space. We are currently investigating models that may yield such
analogous behavior for the black hole case. At the present time, we
consider this to be a potentially viable alternative.

\section{Arguments Against Information Loss} \label{noinfo}

With the possible exception of ``possibility IV'' above, we feel that it is fair
to say that the above alternatives to information loss are neither plausible nor
palatable. Why, then, have people been driven to consider such alternatives?
There are three basic arguments that have been given against information loss.
We now proceed to analyze these arguments.

\medskip

\subsection{Violation of Unitarity} \label{unit}

In scattering theory, the word ``unitarity'' has two completely different
meanings: (i) Conservation of probability. (ii) Evolution from pure states to
pure states.  Failure of (i) would represent a serious breakdown of quantum
theory (and, indeed, of elementary logic). However, it is (ii)---not (i)---that
is being proposed by the semiclassical picture. 

Failure of (ii) would be expected to occur in any situation where the final
``time'' is not a Cauchy surface. Such a failure of unitarity is entirely
innocuous.  For example, we get evolution from a pure state to a mixed state for
a massless Klein-Gordon field in Minkowski spacetime if the final ``time'' is
chosen to be a hyperboloid rather than a hyperplane, as illustrated in Fig.~3.
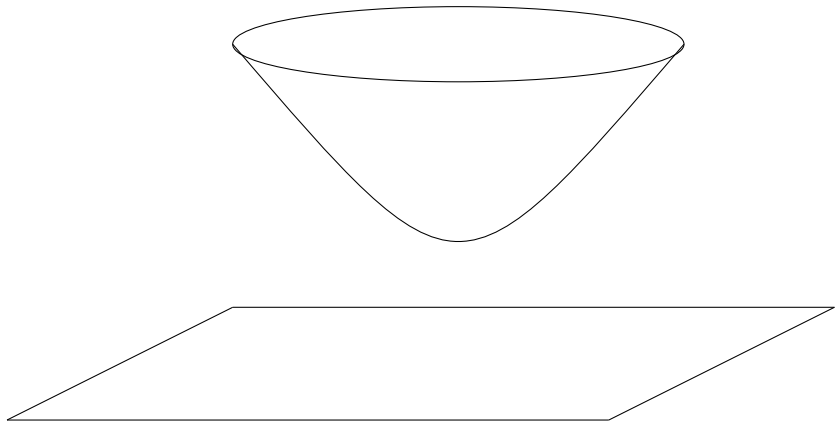
\begin{figure}[ht]
\centering
\begin{tikzpicture}
\draw (0,0) ellipse (-3 and .5);
\draw (-3,0) .. controls (0,-3.5) .. (3,0);
\draw (-6,-5)  -- (2, -5);
\draw (-3,-3.5)  -- (5, -3.5);
\draw (2,-5)  -- (5, -3.5);
\draw (-6,-5)  -- (-3,-3.5);
\end{tikzpicture}
\caption{A hyperboloid in Minkowski spacetime lying to the future of a hyperplane. If we consider
the evolution of a massless quantum field that initially is in a pure state on the hyperplane, it will be 
in a mixed state on the hyperboloid.}
\end{figure}
This is because the state of the quantum field on the hyperboloid is entangled
with the state of the quantum field on the portion of future null infinity that
lies to the past of the cross-section of null infinity corresponding to its
intersection with the hyperboloid. In this case, at the ``time'' represented by
the hyperboloid, information has been ``lost'' to null infinity, leaving the
field on the hyperboloid in a mixed state. 

The situation illustrated in Fig. 3 is not an artificial example but rather
illustrates phenomena that occur around us all of the time. If an atom in your
living room emits a photon, the state of that atom will be in entangled with the
photon. If that photon escapes out the window and is not reflected/absorbed by
clouds or any other intervening matter in the universe, it will be ``lost
forever'' as far as you are concerned. The state of your living room and any
additional portion of the universe that you observe will be mixed.
``Information'' will have been lost. An initial pure state in any experimentally accessible region
around you will have been converted into a mixed state.

The pure state to mixed state evolution predicted by the semiclassical analysis
of black hole evaporation is of an entirely similar character. It is a {\em
prediction} of quantum (field) theory in any situation where the final ``time''
is not a Cauchy surface, not a {\em violation} of quantum theory.

We find it ironic that some researchers who may have been seeking to ``save quantum
mechanics'' by trying to evade the semiclassical arguments that a pure state
evolves to a mixed state are, in fact, effectively attempting to ``destroy
quantum mechanics'' by seeking to modify quantum mechanics (see, e.g., \cite{lp}) 
and/or pursuing
truly drastic alternatives (such as Possibility II of the previous section) that
really are violations of quantum (field) theory in a regime where it should be
valid. We remain firm in our belief in the validity of quantum theory in regimes
away from the Planck scale.

\subsection{Failure of Energy Conservation} \label{en}

It is commonly claimed that any evolution law taking a pure state to a mixed
state would necessarily result in unacceptable violations of energy
conservation. The reference most frequently cited in support of this claim is a
paper of Banks, Peskin, and Susskind \cite{bps}. However, examination of this
reference shows that they considered only a ``Markovian'' type of evolution law,
namely, that given by the Lindblad equation \cite{lin}. It is true that evolution
governed by the Lindblad equation does not conserve energy, although the
violations of energy conservation can be adjusted so as to be negligible at
laboratory scales \cite{uw}. However, Markovian evolution would not be an
appropriate model for black hole evaporation, since the black hole clearly
should retain a ``memory'' of what energy it previously emitted. 

The widespread belief that pure state to mixed state evolution must be
accompanied by violations of energy conservation appears to be closely
associated with a widespread belief that any quantum mechanical decoherence
process requires energy exchange and therefore a failure of conservation of
energy for the system under consideration. The latter belief is true in the case
where the ``environment system'' is taken to be a thermal bath of oscillators
\cite{cl}. However, it is {\it not} true in the case where the environment system
is a ``spin bath'' \cite{ps} where excitation of the degrees of freedom of the
environmental system does not require energy. 

Based on this spin bath idea, Unruh \cite{unr} has provided a simple example of a
quantum mechanical system that interacts with a ``hidden spin system'' in such a
way that an initial pure state of the quantum mechanical system will evolve to a
final mixed state, but {\it exact} energy conservation holds in the process. As
a result of interactions, the quantum mechanical system becomes entangled with the
spin system, but at late times the interactions become negligible, and the spin
system carries no energy.

Unruh's model is not a realistic model for black hole evaporation, and it yields
a relatively small amount of ``information loss.'' Nevertheless, it provides a
proof that there is no problem of principle with maintaining exact energy
conservation in quantum mechanics with an evolution wherein an initial pure state
evolves to a final mixed state. We see absolutely no reason why evolution from a
pure state to a mixed state cannot occur as suggested by the semiclassical
analysis while maintaining exact energy conservation.

\subsection{AdS/CFT} \label{ads}

During the 1980s and most of the 1990s, the main arguments given against
information loss were ``violation of unitarity'' and ``failure of energy
conservation,'' as described in the above subsections. However, since the late
1990s, these arguments have largely been supplanted by the assertion that
evolution from a pure state to a mixed state in the process of black hole
formation and evaporation cannot occur since it would violate the ``AdS/CFT
correspondence.''

The AdS/CFT correspondence is the assertion that quantum gravity (at least on
asymptotically anti-deSitter spacetimes) is ``dual'' to a (non-gravitational)
conformal field theory defined on the boundary of anti-deSitter spacetime. The
one sentence version of AdS/CFT argument against information loss is that since
the conformal field theory---being an ordinary quantum field theory in a fixed
classical spacetime---presumably does not admit pure state to mixed state
evolution, such evolution must also not be possible in quantum gravity,
including when black holes form and evaporate. 

The AdS/CFT correspondence is a conjecture. Our difficulty in assessing the
validity of the AdS/CFT argument against information loss is not so much that
this conjecture has not been {\it proven}, but rather that it has not been {\it
formulated} in sufficient detail and with sufficient precision to make a clear
argument. In particular, relatively little is explicitly known about the
conjectured ``dictionary'' between the ``bulk observables'' in the
asymptotically AdS spacetime and the CFT observables defined in the boundary
theory\footnote{Of course, a key reason why relatively little of the conjectured
dictionary is known is that there is very little, if any, understanding of what
``bulk observables'' are supposed to be in quantum gravity.}. However, as we
shall now explain, the precise nature of this correspondence dictionary is
crucial to using AdS/CFT in arguments against information loss.

One way of formulating the notion of ``information loss'' in black hole
formation and evaporation is the statement that the bulk observables at late
times are not the complete set of bulk observables.  The bulk observables at
late times thereby comprise only part of the independent degrees of freedom of
the bulk system, and the (pure) state of the complete bulk system (which
includes early time observables) is mixed when restricted to the late time
observables. That the bulk state is mixed when restricted to late time
observables is not, by itself, in conflict with the assertion that the complete
set of bulk observables is in 1-1 correspondence with the complete set of CFT
observables. A conflict arises only when one adds additional assumptions about
the correspondence, such as (1) the correspondence is sufficiently ``local in
time'' that the late time bulk observables are in 1-1 correspondence with the
late time CFT observables, and (2) the CFT observables at one time comprise all
of the observables of the CFT system (so that the late time CFT observables are
the complete set of CFT observables). Assumptions of this sort are implicitly
made in all AdS/CFT arguments against information loss.

Assumptions (1) and (2) may appear reasonable, but it is far from clear that
they are valid---even assuming the validity of some version of the AdS/CFT
conjecture. In particular, if the CFT observables can be obtained by taking
limits to the boundary of bulk observables, then (1) would hold automatically
but one would not expect (2) to hold\footnote{For example, one obtains a CFT
by taking a boundary limit of a free scalar field on AdS, but this CFT is a ``generalized 
free field,'' which does not obey assumption (2) \cite{dr}.}. Indeed, if (2) held, then
the bulk observables at all times would be determined by the bulk observables
near infinity at one time, in direct conflict\footnote{This conflict between
predicted behavior in quantum gravity if AdS/CFT together with assumptions (1)
and (2) hold and known classical behavior in general relativity would raise
issues that, in our opinion, are far more significant than the issues raised by
the ``information paradox.'' This remark also applies to the boundary unitarity arguments of \cite{mar}, which can be made independently of the AdSCFT conjecture.} with known classical behavior of general
relativity, as expressed in the ``gluing theorems'' \cite{cip}. On the other hand, if
the CFT observables cannot be obtained as limits to the boundary of bulk
observables, then it is far from obvious that (1) will hold.

It is our hope that the AdS/CFT ideas can be developed further so as to make a
mathematically precise argument regarding information loss in black hole
formation and evaporation. A properly developed argument that AdS/CFT is in
conflict with information loss would necessarily contain an explanation of how
information is regained---and where the semiclassical behavior is violated---not
just an assertion that it must happen somehow or other. With such an argument in
hand, one would have to choose between AdS/CFT and information loss---and one
would be in a position to do so intelligently. At the present time, we see no
necessity of rejecting either alternative.

\section{Conclusions} \label{con}

We have argued that entanglement between the observables outside and inside of a
black hole is a natural and inevitable consequence of the laws of quantum
mechanics and quantum field theory. The proposals to evade or destroy this
entanglement typically require drastic violations of the local laws of physics
in regimes where, {\it a priori}, one would expect them to be valid. We have
also argued that loss of information in black hole formation and evaporation
does not violate any fundamental principles of physics and is not, in any way, a
radical proposal. Thus, our strong inclination is to believe that there is loss
of information in the process of black hole formation and evaporation. 

\bigskip
\noindent
{\bf Acknowledgements}

The research of W.G.U. was supported by the Natural Sciences and Engineering Research Council of Canada, the Canadian Institute for Advanced Research, the Perimeter Institute, and the
Templeton Foundation. The research of R.M.W was supported by NSF grant PHY 15-05124 to the University of Chicago. We thank numerous colleagues---far too many to name individually here---for many extremely stimulating discussions on this topic during the past 40 years. Of particular note were numerous interactions that we both had at the KITP workshop ``Black Holes: Complementarity, Fuzz, or Fire?" in August, 2013. We also thank Don Marolf for several comments on a draft of this manuscript.

\end{document}